\documentclass{article}
\usepackage{epsf,graphics,amsmath,epsfig,latexsym,amssymb,cite,graphicx,a4}
\newlength{\indentedwidth} \newdimen\mathindent
\indentedwidth=\mathindent

\usepackage{epsf,graphics}
\DeclareMathAlphabet{\mathpzc}{OT1}{pzc}{m}{it}
\begin{document}
\vskip 0.5cm
\begin{center}
{\Large \bf Generalised Perk--Schultz models: solutions of the Yang-Baxter equation associated with quantised
orthosymplectic superalgebras}
\end{center}
\vskip 0.8cm
\centerline{M. Mehta,% 
\footnote{\tt maitha7@gmail.com} K.A. Dancer,%
\footnote{\tt dancer@maths.uq.edu.au} M.D. Gould %
 and J. Links%
\footnote{\tt jrl@maths.uq.edu.au}}
\vskip 0.9cm \centerline{\sl\small Centre for Mathematical Physics,
School of Physical Sciences, } \centerline{\sl\small The University of
Queensland, Brisbane 4072, Australia.}
\vskip 0.9cm

\def\uq{$U_{q}[osp(m|n)]$}
\def\gl{$gl(m|n)$}
\def\z2{$\mathbb{Z}_{2}$-graded}
\def\rh{\check{R}}
\def\q-{(q-q^{-1})}
\def\v2{$V(2\d_1 )$}
\def\qq{$(q+q^{-1})$}
\def\qr{(q+q^{-1})}
\def\vi{\tilde{v}^i}
\def\vm{\tilde{v}^{\mu}}
\def\vs{\tilde{v}^s}
\def\p2{$\pro{2\d_{1}}$}
\def\pp1{\pro{2\d_{1}}}
\def\bg{\s}
\def\s{\sigma}
\def\d{\delta}
\def\p{\psi}
\def\x{\xi}
\def\r{\rho}

\newcommand{\ve}[1]{\ensuremath{\varepsilon_{#1}}}
\newcommand{\bk}[2]{ \ensuremath{| #1 \rangle\langle #2 |}}
\newcommand{\pro}[1]{\ensuremath{ \mathbb{P}_{#1}}}
\newcommand{\vc}[6]{\ensuremath{\sum^{#1}_{#2}{#3}q^{#4}w_{#5}\otimes w_{#6}}}
\newcommand{\vl}[8]{\ensuremath{\sum^{#1}_{#2}{#3}q^{#4}E^{#5}_{#6}\otimes E^{#7}_{#8}}}
\newcommand{\bv}[3]{\ensuremath{q^{#1} w_{#2}\otimes w_{#3}}}
\newcommand{\ev}[2]{\ensuremath{w_{#1}\otimes w_{#2}}}
\newcommand{\su}[2]{\ensuremath{\sum^{#1}_{#2}}}
\newcommand{\qs}[4]{\ensuremath{\frac{q^{#1}+q^{#2}}{q^{#3}+q^{#4}}}}
\newcommand{\ip}[2]{\ensuremath{\langle v_{#1}, v_{#2}\rangle}}
\newcommand{\eg}[2]{\ensuremath{E^{#1}_{#2}}}
\newcommand{\de}[2]{\ensuremath{\d^{#1}_{#2}}}
\newcommand{\vq}[8]{\ensuremath{\sum^{#1}_{#2}(-1)^{#3}q^{#4}E^{#5}_{#6}\otimes E^{#7}_{#8}}}
\newcommand{\vv}[7]{\ensuremath{\sum^{#1}_{#2}(-1)^{#3}E^{#4}_{#5}\otimes E^{#6}_{#7}}}
\newcommand{\sv}[6]{\ensuremath{\sum^{#1}_{#2}E^{#3}_{#4}\otimes E^{#5}_{#6}}}
\newcommand{\es}[6]{\ensuremath{\sum^{#1}_{#2}E^{#3}_{#4}\otimes \hat{\s}^{#5}_{#6}}}
\newcommand{\suv}[4]{\ensuremath{\sum^{#1}_{#2}(E^{#3}_{#3}\otimes E^{#4}_{#4}+E^{#4}_{#4}\otimes E^{#3}_{#3})}}  
\newcommand{\eso}[7]{\ensuremath{\sum^{#1}_{#2}(-1)^{#3}E^{#4}_{#5}\otimes \hat{\s}^{#6}_{#7}}}

\begin{abstract} 

The Perk--Schultz model may be 
expressed in terms of the solution of the Yang--Baxter equation associated with 
the fundamental representation of the untwisted affine extension 
of the general linear quantum superalgebra $U_q[sl(m|n)]$, with a multiparametric
co-product action as given by Reshetikhin.
Here we present analogous explicit expressions for solutions of the Yang-Baxter equation 
associated with the fundamental representations of the twisted and untwisted affine 
extensions of the orthosymplectic quantum superalgebras \uq. 
In this manner we obtain generalisations of the Perk--Schultz model.  
\end{abstract} 

%Classifications:    \\
Keywords:

%%%%%%%%%%%%%%%%%%%%%%%%%%%%%%%%%%%%%%%%%%%%
\begin{section}{Introduction} 

The Perk--Schultz model \cite{schultz,perk} is well known to be 
exactly solvable \cite{lopes}. For fixed $d>1$, the model is defined on a square lattice where each 
edge can occupy one of $d$ states. In addition to the spectral parameter 
the model depends on $1+d(d-1)/2$ continuous variables, and $d$ 
discrete variables which have value $\pm 1$. 
One method to formulate the model and obtain the exact solution 
is through the $R$-matrix associated with the fundamental representation of the 
quantised untwisted affine general linear superalgebra $U_q[sl(m|n)^{(1)}]$ \cite{okado}. The exact solution follows 
from the fact that the $R$-matrix satisfies the Yang--Baxter equation.    
In this setting, the continuous variables are given by the deformation parameter $q$, as well 
as $d(d-1)/2$ variables associated with the Reshetikhin twist \cite{okado,reshetikhin} on the co-algebra structure.
The discrete variables are associated with the $\mathbb{Z}_{2}$-grading of the $d$-dimensional vector
space which affords the representation of the $U_q[sl(m|n)^{(1)}]$ superalgebra, where 
$m+n=d$. 

Here we report the extension of this result to the case of the 
quantised untwisted affine superalgebra $U_q[osp(m|n)^{(1)}]$ and the twisted case 
$U_q[sl(m|n)^{(2)}]$ where $n=2k$ is even in both instances.  
%Solutions of the Yang-Baxter equation associated with quantised
%deformations of 
%twisted and untwisted general linear, orthogonal and symplectic Lie
%algebras in their
%fundamental representations have long been known. The extension
%of these results
%to supersymmetric (or more precisely $\mathbb{Z}_{2}$-graded) Lie
%algebras has been
%incomplete. The untwisted supersymmetric general linear (i.e.
%$U_{q}(gl(m|n)^{(1)})$)
%solutions fall into the class of Perk-Schultz models \cite{}. The
%corresponding results for
%the untwisted orthosymplectic case $U_{q}(osp(m|n)^{(1)})$ and twisted
%general linear
%superalgebras $U_{q}(gl(m|n)^{(2)})$ are  not known apart from 
%some particular cases \cite{DGLZ95, S90, KR89}. 
%
%Formal expressions
%exist \cite{} but
%explicit formulae in terms of the matrix elements in the tensor product
%space on which the
%solution acts are lacking. Here we will present the spectral parameter
%dependent
%$R$-matrices, which solve the Yang-Baxter equation, associated with the
%latter quantised
%superalgebras. 
A representation theoretic approach is adopted to find 
$R$-matrices satisfying the $\mathbb{Z}_{2}$-graded 
Yang--Baxter equation (YBE)
\begin{equation*}
R_{12}(z)R_{13}(zw)R_{23}(w)=R_{23}(w)R_{13}(zw)R_{12}(z)
\end{equation*}
where $R(z)\in {\rm End} (V(\d_{1})\otimes V(\d_{1}))$ 
and $V(\d_{1})$ is the $(m+n)$-dimensional 
space for the vector representation of \uq \,of highest weight $\d_1$. 
The multiplication on the tensor product space is \z2 (see equation \ref{tp} in the following section).
The construction of $R$-matrices satisfying the \z2 YBE for the general case
$V(\lambda_{a})\otimes V(\lambda_{b})$ (where $\lambda_a,\,\lambda_b$ are the highest
weights of the modules) has been delineated in \cite{DGLZ95,GZ00}. 
In those works, the solutions are presented in general terms 
as a linear combination of 
elementary intertwiners, where the co-efficients are determined through 
tensor product graph methods. However, to have fully complete expressions 
it is necessary to determine also the form of the \uq \, invariant intertwiners 
which project out the submodules in the tensor product decomposition. 
Here,
we will explicitly formulate $R$-matrices for the case $V(\d_{1})\otimes
V(\d_{1})$ for \uq , in both the twisted and untwisted cases by explicitly
computing the elementary intertwiners.
We mention that formal expressions for the solutions of the Yang--Baxter equation
associated with fundamental representations of superalgebras are given
in \cite{bazhanov}, which may also be used to determine explicit expressions
for the $R$-matrices (e.g.\cite{galleas}). 
An alternative approach is to use the Lax operator method
as described in \cite{karen1,karen2}. 

Once the explicit $R$-matrices have been obtained, we will introduce the 
Reshetikhin twist \cite{reshetikhin} in order to generate more general $R$-matrices 
with additional free parameters. 
These results can be used to obtain classes of integrable 
Hamiltonians 
describing systems of interacting fermions, with potential 
applications in condensed
matter systems (cf. \cite{flr}). 
\end{section}

%%%%%%%%%%%%%%%%%%%%%%%%%%%%%%%%%%%%%%%%%
\section{The quantised orthosymplectic superalgebra $U_q[osp(m|n)]$}

\noindent The quantum superalgebra $U_q[osp(m|n)]$ is a
$q$-deformation of the classical orthosymplectic superalgebra.  A
brief explanation of $U_q[osp(m|n)]$ is given below, with more
details to be
found in \cite{karen1}.  Throughout we use $n=2k$ and $l = \lfloor \frac{m}{2} 
\rfloor$, so $m=2l$ or $m=2l+1$.

First we need to define the notation.  The grading of $a$ is
denoted by $[a]$, where

\begin{equation*}
[a] =
   \begin{cases}
     0, \qquad \; a=i, & 1 \leq i \leq m, \\ 1, \qquad \; a = \mu, & 1
     \leq \mu \leq n.
   \end {cases}
\end{equation*}

\noindent We also use the symbols $\overline{a}$ and $\xi_a$, which are 
defined by:

\begin{equation*}
\overline{a}=
  \begin{cases}
     m + 1 - a, & [a]=0, \\ n + 1 - a, & [a]=1,
  \end {cases}
\qquad \text{and} \quad \xi_{a} =
  \begin{cases}
     1, & [a] = 0, \\ (-1)^a, & [a]=1.
  \end{cases}
\end{equation*}

As a weight system for $U_q[osp(m|n)]$ we take the set $\{
\varepsilon_i,\; 1 \leq i \leq m \} \cup \{ \delta_\mu,\; 1 \leq \mu
\leq n \}$, where \mbox{$\varepsilon_{\overline{i}} = -
\varepsilon_i$} and \mbox{$\delta_{\overline{\mu}} = - \delta_\mu$}.
Conveniently, when $m=2l+1$ this implies $\varepsilon_{l+1} =
-\varepsilon_{l+1} = 0$.  Acting on these weights, we have the
invariant bilinear form defined by:

\begin{equation*}
(\varepsilon_i, \varepsilon_j) = \delta^i_j, \quad (\delta_\mu,
\delta_\nu) = -\delta^\mu_\nu, \quad (\varepsilon_i, \delta_\mu) = 0,
\qquad 1 \leq i, j \leq l, \quad 1 \leq \mu, \nu \leq k.
\end{equation*}

The even positive roots of $U_q[osp(m|n)]$ are composed
entirely of the usual positive roots of $o(m)$ together with those of
$sp(n)$, namely:

\begin{alignat*}{3}
&\varepsilon_i \pm \varepsilon_j, & \qquad & 1 \leq i < j \leq l, \\
&\varepsilon_i, && 1 \leq i \leq l &&\quad \text{when } m=2l+1, \\
&\delta_\mu + \delta_\nu, && 1 \leq \mu,\nu \leq k, \\ &\delta_\mu -
\delta_\nu, && 1 \leq \mu < \nu \leq k.
\end{alignat*}

\noindent The root system also contains a set of odd positive roots,
which are:

\begin{equation*}
\delta_\mu + \varepsilon_i, \qquad 1 \leq \mu \leq k,\;1 \leq i \leq
m.
\hspace{2cm}
\end{equation*}

\noindent Throughout this paper we choose to use the following set of
simple roots:

\begin{align*}
&\alpha_i = \varepsilon_i - \varepsilon_{i+1}, \hspace{11mm} 1 \leq i
  < l, \notag \\ &\alpha_l =
\begin{cases} 
\varepsilon_{l} + \varepsilon_{l-1},\quad & m=2l, \\ \varepsilon_l,
&m=2l+1,
\end{cases} \notag \\
&\alpha_\mu = \delta_\mu - \delta_{\mu+1}, \hspace{9mm} 1 \leq \mu <
k,\notag\\ &\alpha_s = \delta_k - \varepsilon_1.
\end{align*}

\noindent Note this choice is only valid for $m >2$.  Also observe that the 
graded half-sum of positive roots is given by:

\begin{equation*}
\rho = \frac{1}{2} \sum_{i=1}^l (m-2i) \varepsilon_i + \frac{1}{2}
\sum_{\mu=1}^k (n-m+2-2\mu) \delta_\mu.
\end{equation*}

In $U_q[osp(m|n)]$ the graded commutator is realised by

\begin{equation*}
[A,B] = AB - (-1)^{[A][B]} BA
\end{equation*}
 
\noindent and tensor product multiplication is given by

\begin{equation} \label{tp}
(A \otimes B) (C \otimes D) = (-1)^{[B][C]} (AC \otimes BD).
\end{equation}

\noindent Using these conventions, the quantum superalgebra $U_q[osp(m|n)]$ is 
generated by simple generators $e_a, f_a, h_a$ subject to relations including:

\begin{align*}
&[h_a, e_b] = (\alpha_a, \alpha_b) e_b, \quad [h_a, f_b] = -
(\alpha_a, \alpha_b) f_b,  \quad [h_a, h_b] = 0, \\
&[e_a, f_b] = \delta^a_b \frac{(q^{h_a} - q^{-h_a})}{(q - q^{-1})},
\quad [e_s, e_s] = [f_s,f_s]=0.
\end{align*}

We remark that $U_q[osp(m|n)]$ has the structure of a quasi-triangular
Hopf superalgebra. In particular, there is a superalgebra homomorphism known 
as the \textit{coproduct},
$\Delta: U_q[osp(m|n)] \rightarrow U_q[osp(m|n)]^{\otimes
2}$, which is defined on the simple generators by:

\begin{align*}
&\Delta (e_a) = q^{\frac{1}{2}h_a} \otimes e_a + e_a \otimes
  q^{-\frac{1}{2} h_a}, \notag \\ &\Delta (f_a) = q^{\frac{1}{2}h_a}
  \otimes f_a + f_a \otimes q^{-\frac{1}{2} h_a}, \\ 
&\Delta (q^{\pm \frac{1}{2}h_a}) = q^{\pm \frac{1}{2}h_a} \otimes q^{\pm
  \frac{1}{2}h_a}.
\end{align*}

\noindent Also, $U_q[osp(m|n)]$ contains a \textit{universal $R$-matrix} 
 which satisfies, among other properties, the {\it Yang--Baxter equation}: 

\begin{equation*}
\mathcal{R}_{12} \mathcal{R}_{13} \mathcal{R}_{23} = \mathcal{R}_{23}
  \mathcal{R}_{13} \mathcal{R}_{12}.
\end{equation*}

\noindent Here $\mathcal{R}_{ab}$ represents a copy of $\mathcal{R}$
acting on the $a$ and $b$ components respectively of $U \otimes U
\otimes U$, where each $U$ is a copy of the quantum superalgebra
$U_q[osp(m|n)]$.

 Now let $\text{End} \; V$ be the space of endomorphisms of
$V$, an $(m+n)$-dimensional vector space.  Then the irreducible
\textit{vector representation} $\pi: U_q[osp(m|n)] \rightarrow
\text{End} \; V$ acts on the $U_q[osp(m|n)]$ generators as 
given in Table \ref{vector}, where $E^a_b$ is the elementary matrix with a 1 
in the $(a,b)$ position and zeroes elsewhere. 

%Table should be as close to the definition of the vector representation as possible
\begin{table}[ht] 
\caption{The action of the vector representation $\pi$ on the simple 
generators of $U_q[osp(m|n)]$} 
\label{vector}
\centering
\begin{tabular}{|l|l|l|l|}\hline
\multicolumn{1}{|c|}{$\alpha_a$}& \multicolumn{1}{c|}{$\pi(e_a)$} & 
  \multicolumn{1}{c|}{$\pi(f_a)$} & \multicolumn{1}{c|}{$\pi(h_a)$}\\ \hline 
$\alpha_i, 1 \leq i < l$ &
  $E^i_{i+1} - E^{\overline{i+1}}_{\overline{i}}$&
  $E_i^{i+1} - E_{\overline{i+1}}^{\overline{i}}$&
  $E^i_i - E^{\overline{i}}_{\overline{i}} - E^{i+1}_{i+1}
             + E^{\overline{i+1}}_{\overline{i+1}}$\\ 
$\alpha_l, \, m=2l$ &
  $E^{l-1}_{\overline{l}} - E^l_{\overline{l-1}}$ &
  $E_{l-1}^{\overline{l}} - E_l^{\overline{l-1}}$&
  $E^{l-1}_{l-1} + E^l_l - E^{\overline{l-1}}_{\overline{l-1}}
             - E^{\overline{l}}_{\overline{l}}$ \\
$\alpha_l, \, m=2l+1$ 
  &$E^l_{l+1} - E^{l+1}_{\overline{l}} $ &
  $E_l^{l+1} - E_{l+1}^{\overline{l}}$&
  $E^l_l - E^{\overline{l}}_{\overline{l}} $\\
$\alpha_\mu, 1 \leq \mu <k$ & 
  $E^\mu_{\mu+1} + E^{\overline{\mu+1}}_{\overline{\mu}}$&
  $E_\mu^{\mu+1} + E_{\overline{\mu+1}}^{\overline{\mu}}$&
  $E^{\mu+1}_{\mu+1} - E^{\overline{\mu+1}}_{\overline{\mu+1}}
               - E^\mu_\mu + E^{\overline{\mu}}_{\overline{\mu}}$ \\
$\alpha_s$ &
  $E^{\mu=k}_{i=1} + (-1)^k E^{\overline{i=1}}_{\overline{\mu=k}}$&
  $- E^{i=1}_{\mu=k} + (-1)^k E_{\overline{i=1}}^{\overline{\mu=k}}$&
  $- E^{i=1}_{i=1} + E^{\overline{i}=\overline{1}}_{\overline{i}=\overline{1}} 
   -  E^{\mu=k}_{\mu=k} + E^{\overline{\mu}=\overline{k}}_
           {\overline{\mu}=\overline{k}}$ \\ \hline
\end{tabular}
\end{table}

The solutions to the Yang--Baxter equation in a given representation of $U_q[osp(m|n)]$ can sometimes be extended to solutions of the spectral parameter dependent Yang--Baxter equation

$$R_{12}(z) R_{13}(zw) R_{23}(w) = R_{23}(w) R_{13}(zw) R_{12}(z)$$

\noindent in the affine extensions $U_q[osp(m|n)^{(1)}]$ and 
$U_q[gl(m|n)^{(2)}]$.  In the following sections we construct such solutions 
for the case of the vector representation.

%%%%%%%%%%%%%%%%%%%%%%%%%%%%%%%%%%%%%%%%%%%%%%5

\begin{section}{Determination of  the $R$-matrices}

The tensor product of the vector module with itself decomposes into 
$U_q[osp(m|n)]$ modules according to

\begin{equation*}
V(\d_1) \otimes V(\d_1) =V(2\d_1)\oplus V(\d_1 +\d_2) \oplus V(\dot{0})
\end{equation*}

\noindent except in the case $m=n$, in which case the last two irreducible 
modules combine to form an indecomposable $V$.
Let

\begin{eqnarray*}
\mathbb{P}_{V}= & \left\{ \begin{array}{ll} 
   V(\d_1 +\d_2) \oplus V(\dot{0}) & \hbox{ for } m\neq n\\
   V  \hbox{ indecomposable }              & \hbox{ for } m=n.
			  \end{array} \right. 
\end{eqnarray*}

\noindent Then we have a resolution of the identity as follows:

\begin{equation*}
I=\mathbb{P}_{2\d_{1}}+\mathbb{P}_{V}.
\end{equation*}

Define $\rh(z)=PR(z)$ where $P=\sum_{a,b} (-1)^{[b]}e^a_b\otimes e^b_a$ 
is the graded permutation operator.  Then the Yang-Baxter 
equation may be rewritten as

$$\rh_{12}(z)\rh_{23}(zw)\rh_{12}(w)=\rh_{23}(w)\rh_{12}(zw)\rh_{23}(z).$$  

\noindent {}From \cite{DGLZ95,GZ00} it is known that

\begin{equation}
\rh=\sum_{a}\rho_{a}(z)\mathbb{P}_a  \label{rmat}
\end{equation}

\noindent where $\mathbb{P}_{a}$ denotes the $U_q[osp(m|n)]$ invariant 
projection operator onto the submodule $V(a)$.  The co-efficients 
$\rho_{a}(z)$ are determined using

\begin{equation}
\rho_{a}(z)=
\left\langle\frac{C(a^{\prime})-C(a)}{2}  \right\rangle_{\epsilon_a 
 \epsilon_{a^{\prime}}} \rho_{a^{\prime}}(z) \label{rho}
\end{equation}

\noindent where

\begin{equation*}
\langle x\rangle_{\pm}=\frac{1\pm zq^x}{z\pm q^x}
\end{equation*}

\noindent provided the weights $a,\, a^{\prime}$ label adjacent vertices in 
the {\it tensor product graph} \cite{DGLZ95,GZ00}. Here $C(a)$ denotes the eigenvalue of 
the second order Casimir invariant on $V(a)$ and $\epsilon_a$ the parity of 
the vertex associated with $a$.  For $U_q[osp(m|n)^{(1)}]$, the tensor 
product graph is depicted in Figure \ref{untwisted}
while the tensor product graph for $U_q[gl(m|n)^{(2)}]$ is given in Figure \ref{twisted}.  

\begin{figure}[h]
\begin{center}
  \scalebox{1}{\input{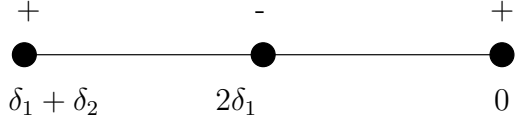}}
\end{center}\caption{The untwisted tensor product graph \label{untwisted}}
\end{figure}

\begin{figure}[h] 
\begin{center}
\scalebox{1}{\input{ttpg.pstex_t}}
\end{center}\caption{The twisted tensor product graph \label{twisted}}
\end{figure}

Let $\psi$ denote the (unnormalised)
basis vector for the identity module $V(\dot{0})$. Explicitly 
\begin{equation*}
\psi =\psi_0 +\psi_1
\end{equation*}
where 
$$\psi_0 = \vc{m}{i=1}{}{-(\rho,\ve{i})}{i}{\overline{i}}$$ 
and
$$\psi_1 = \vc{n}{\mu =1}{-(1)^{\mu}}{-(\rho,\d_{\mu})}{\mu}{\overline{\mu}}.$$

>From equations (\ref{rmat}) and (\ref{rho}), 
we find that for $U_q[osp(m|n)^{(1)}]$ the required $R$-matrix is 

\begin{equation}
\rh(z)=\mathbb{P}_{2\d_1}+\frac{1-zq^{-2}}{z-q^{-2}}\mathbb{P}_{\d_1 +\d_2}+  
\left(\frac{1-zq^{m-n-2}}{z-q^{m-n-2}}\right)\mathbb{P}_{0}   \label{rh}
\end{equation}

\noindent where 

\begin{equation*}
\pro{0}=\frac{1}{1-[n+1-m]_q}\bk{\psi}{\psi}
\end{equation*}

\noindent and $[k]_q =\frac{q^k -q^{-k}}{q-q^{-1}}$.
For $U_q[gl(m|n)^{(2)}]$ we obtain the analogous result 

\begin{equation}
\rh=\mathbb{P}_{2\d_1}+\frac{1-zq^{-2}}{z-q^{-2}}\mathbb{P}_{\d_1 +\d_2}+
\left(\frac{1+zq^{m-n}}{z+q^{m-n}}\right)
\left(\frac{1-zq^{-2}}{z-q^{-2}}\right)\mathbb{P}_{0}.   \label{rh1}
\end{equation}

\noindent Note that in equations (\ref{rh},\ref{rh1}) \pro{0} is not defined 
for $m=n$. To avoid having to make separate calculations, define 

\begin{eqnarray*}
Q & = & \frac{\q- q^{-1}}{(q^{m-n-2}+1)}\bk{\psi}{\psi}\\
  & = & (1-q^{n-m})\pro{o}.
\end{eqnarray*}

\noindent Then $\rh(z)$ can be written (and renormalised) as

\begin{eqnarray*}
\rh(z) & = & \frac{z-q^{-2}}{1-zq^{-2}}{\pp1} + \pro{\d_1 +\d_2} 
+\left(\frac{z-q^{-2}}{1-zq^{-2}}\right)
\left(\frac{1-zq^{m-n-2}}{z-q^{m-n-2}}\right) \pro{0}\\
    & = & \frac{(1+q^{-2})(z-1)}{1-zq^{-2}}{\pp1} +I 
+\frac{(z^2-1)}{(zq^{-2}-1)(zq^{n-m+2}-1)} Q
\end{eqnarray*}

\noindent for $U_q[osp(m|n)^{(1)}]$ and 
\begin{eqnarray*}
\rh(z) & = & \frac{z-q^{-2}}{1-zq^{-2}}{\pp1} + \pro{\d_1 +\d_2} +\frac{1+zq^{m-n}}{
z+q^{m-n}} \pro{0}\\
    & = & \frac{(1+q^{-2})(z-1)}{1-zq^{-2}}{\pp1} +I +\frac{(z-1)q^{m-n}}{z+q^{m-
n}} Q
\end{eqnarray*}
for $U_q[gl(m|n)^{(2)}]$. 

In order to obtain explicit expressions for the $R$-matrices, it 
remains to determine the operator ${\pp1}$. First we find the following 
orthogonal basis vectors for $V(2\d_1 )$:

\begin{eqnarray*} 
\bv{-1/2}{i}{j}-\bv{1/2}{j}{i}, & w_{\mu}\otimes w_{\mu},\\
\bv{-1/2}{\mu}{\nu}+\bv{1/2}{\nu}{\mu}, & \bv{1/2}{i}{\mu}-\bv{-1/2}{\mu}{i}, 
\end{eqnarray*}

\noindent where $1\leq\mu < \nu\neq\overline{\mu}\leq n$, 
and $1\leq i<j\neq\overline{i}\leq n.$
The zero weight vectors are given by the following:

\

\( \begin{array}{llll}
v_i     & =\ev{i}{\overline{i}} -\ev{\overline{i}}{i} -\bv{-1}{i+1}
          {\overline{i+1}} +\bv{}{\overline{i+1}}{i+1}, 
          \hspace{19.5mm} 1 \leq i <l\\
v_s     & =\bv{-1}{1}{\overline{1}} -\bv{}{\overline{1}}{1} +(-1)^k 
          (\bv{-1}{k}{\overline{k}} +\bv{}{\overline{k}}{k})\\ 
v_{\mu} & =(-1)^{\mu}(\bv{-1}{\mu}{\overline{\mu}} +\bv{}{\overline{\mu}}{\mu} 
          +\ev{\mu +1}{\overline{\mu +1}}+\ev{\overline{\mu +1}}{\mu+1} ), 
          \;\; 1 \leq \mu < k \\
v_l     & =\ev{l}{\overline{l}} -\ev{\overline{l}}{l} + 
  \left\{  \begin{array}{ll}  0,   &  m=2l\\
                        (q^{1/2}-q^{-1/2})\ev{l+1}{l+1}, &  m=2l+1
             \end{array}  \right. 
\end{array} \) \newline

\noindent These, however, are not orthogonal.  Instead, we complete an 
orthogonal dual basis for $V(2 \d_1)$ with the following orthogonal 
zero-weight dual vectors:

\begin{alignat*}{2}
v^i     & =  \vi +\frac{D_{l-i}[k]_q}{\qr D_{l-k}}\Omega, 
          \qquad &&1 \leq i \leq l, \\
v^{\mu} & =  \vm +\frac{[\mu]D_l}{\qr D_{l-k}}\Omega, 
          &&1 \leq \mu < k, \\
v^s     & =  \frac{[k]D_l}{\qr D_{l-k}} \Omega, &&
\end{alignat*}

\noindent where 

\begin{eqnarray*}  
\tilde{v}^i     & = & \frac{1}{(q+q^{-1} )D_l } \left\{ [i]_q \sum^{l}_{j\geq i} D_{l-j}v_j +D_{l-i}\sum_{j<i} [j]_q
v_j \right\},\\
\tilde{v}^{\mu} & = & \frac{-1}{ (q+q^{-1} )[k]_q } \left\{ [\mu]_q \su{k-1}{\nu\geq\mu}[k-\nu]_q v_{\nu}+[k-\mu]_q
\su{}{\nu <\mu}[\nu]_q
v_{\nu} \right\}
\end{eqnarray*}

and 

\begin{equation}
D_x = \left\{ \begin{array}{ll}                      
                \qs{x-1}{x-l}{}{-1}, & m=2l\\
                \qs{x-1/2}{1/2-x}{1/2}{-1/2}, & m=2l+1
                \end{array}  \right\}  \nonumber         
= \qs{x+\frac{m}{2}-l-1}{l+1-x-\frac{m}{2}}{\frac{m}{2}-l-1}{l+1-\frac{m}{2}}.
\end{equation}

It is convenient at this point to introduce the braid generator, $\bg$:

\begin{eqnarray*}
\bg & = & q^{-1}\check{R}(0)\\
    & = & \qr\pp1 -qI + \frac{\q- }{q^{m-n-2}+1}\bk{\psi}{\psi}.
\end{eqnarray*}

\noindent Note that $\check{R}(0)$ is the same for both $U_q[osp(m|n)]$ and 
$U_q[gl(m|n)^{(2)}]$. After calculating $\pp1$ and \bk{\psi}{\psi}, 
we find this explicit expression for the braid generator $\bg$:

\begin{multline*}
\bg = -\vv{}{a\neq b,\bar{b}}{[b]}{a}{b}{b}{a} -\vq{}{a}{[a]}{(\ve{a}, \ve{a} )}{a}{a}{a}{a}\\
+ \q-\left\{\su{l}{i=1}\left [ \vl{}{i\leq j\leq\bar{i}}{}{-(\rho, \ve{i}+ \ve{j})}{i}{j} 
{\bar{i}}{\bar{j}} +\vl{}{i<j<\bar{i}}{}{-(\rho,\ve{i}+ \ve{j})}{j}{i}{\bar{j}}{\bar{i}}\right ]\right .\\
-\vq{}{\mu\leq\nu\leq\bar{\mu}}{\mu +\nu}{-(\rho ,\d_{\mu} +\d_{\nu})}{\mu}{\nu}{\bar{\mu}}{\bar{\nu}}
  - \vq{}{\mu <\nu <\bar{\mu}}{\mu +\nu }{-(\rho ,\d_{\mu} +\d_{\nu})}{\nu}{\mu}{\bar{\nu}}{\bar{\mu}}\\ 
 +\su{k}{\mu =1} \su{m}{i=1}(-1)^{\mu}q^{-(\rho, \ve{i} +\d_{\mu})}(E^{i}_{\mu} \otimes E^{\bar{i}} 
_{\bar{\mu}}+E^{\mu}_i \otimes E^{\bar{\mu}}_{\bar{i}})\\
 -\q- \left\{\su{m}{i<j} E^i_i \otimes E^j_j +\su{n}{\mu <\nu} E^\mu_\mu \otimes E^\nu_\nu +\su{m}{i=1}\su{k}{\mu
=1} (E^i_i \otimes E^{\bar{\mu}}_{\bar{\mu}} +E^\mu_\mu \otimes E^i_i )\right\}\\
-\su{l}{i=1} (qE^i_{\bar{i}}\otimes E^{\bar{i}}_i + q^{-1}E^{\bar{i}}_i \otimes E^i_{\bar{i}}) + \su{k}{\mu =1}
(q^{-1}E^\mu_{\bar{\mu}}\otimes E^{\bar{\mu}}_\mu +qE^{\bar{\mu}}_\mu \otimes E^\mu_{\bar{\mu}})
\end{multline*}

Recall  the relation $R(z)=P\check{R}(z)$.
If we substitute in the previous equation and simplify, 
we obtain an expression for the
$R$-matrices in the zero spectral parameter limit 
which we will denote by $R^{\prime}$:

\begin{multline*}
q^{-1}R^{\prime} = -\sv{}{a\neq b, \bar{b}}{b}{b}{a}{a} - \vl{}{a}{}{(\ve{a} ,\ve{a} )}{a}{a}{a}{a}\\
-q^{-1}\suv{l}{i=1}{i}{\bar{i}} - q\suv{k}{\mu=1}{\mu}{\bar{\mu}}\\
-\q-\left\{\su{m}{i>j}E^i_j \otimes\hat{\s}^j_i - \su{n}{\mu >\nu}
E^\mu_\nu \otimes\hat{\s}^\nu_\mu+ \su{m}{i=1}\su{k}{\mu=1}(E^{\bar{\mu}}_i \otimes\hat{\s}^i_\mu - E^i_\mu \otimes\hat{\s}^\mu_i)
\right\}
\end{multline*}

\noindent where

\begin{equation*}
\hat{\s}^a_b =E^a_b- (-1)^{[a]([a]+[b])}\xi_a\xi_b q^{(\rho,\ve{b}
-\ve{a})}E^{\bar{b}}_{\bar{a}}
\end{equation*}

\noindent and

\begin{equation*}
\hat{\s}^a_a =q^{1/2(\ve{a} ,\ve{a} )}E^a_a - q^{-1/2(\ve{a} , \ve{a} )}
E^{\bar{a}}_{\bar{a}}.
\end{equation*}

\noindent This equation simplifies further to give 

\begin{multline*}
q^{-1}R^{\prime} = -I -(q^{1/2}-q^{-1/2})\su{}{a}(-1)^{[a]} E^a_a \otimes\hat{\s}^a_a \\
-\q- \left\{ \su{m}{i>j} E^i_j \otimes\hat{\s}^j_i -  \su{n}{\mu >\nu}
E^\mu_\nu \otimes\hat{\s}^\nu_\mu 
 + \su{k}{\mu =1}\su{m}{i=1}(E^{\bar{\mu}}_i \otimes \hat{\s}^i_{\bar{\mu}} -
E^i_\mu \otimes\hat{\s}^\mu_i)\right\}.
\end{multline*}

We now rewrite $\check{R}(z)$ for $U_q[osp(m|n)^{(1)}]$ in terms of the braid 
generator $\bg$.

\begin{equation*}
\rh(z) =\frac{1}{(q-q^{-1}z)}\left\{(z-1)\bg + \q- zI -\frac{\q- z(z-1)}
{(z -q^{m-n-2})}\bk{\p}{\p}\right\}. \label{rbg}
\end{equation*}

\noindent Using equation (\ref{rbg}), we can determine the normalized 
$R$-matrices as follows

\begin{equation*}
R(z) =\frac{1}{(q-q^{-1}z)}\left\{(z-1)q^{-1}R^{\prime} + \q- zP -\frac{\q- 
z(z-1)}{(z -q^{m-n-2})} P\bk{\p}{\p}\right\}.  
\end{equation*}

\noindent Explicit calculation gives the following expansion for $R(z)$ in 
the untwisted case:

\begin{multline*}
R(z) = \frac{\q- zP}{(q-q^{-1}z)} - \frac{\q-z(z-1)}{(q-q^{-1}z)(z-q^{m-n-2})}
  \vl{}{a,b}{(-1)^{[a][b]}\xi_a\xi_b}
  {(\r,\ve{a}-\ve{b})}{a}{b}{\bar{a}}{\bar{b}}\\
-\frac{(z-1)}{(q-q^{-1}z)}\left\{ I + (q^{1/2}-q^{-1/2})
  \eso{}{a}{[a]}{a}{a}{a}{a} + \q- \sum_{\ve{a} < \ve{b}} (-1)^{[b]} 
  E^a_b \otimes \hat{\sigma}^b_a \right\}.
\end{multline*} 

\noindent Similarly, for $U_q[gl(m|n)^{(2)}]$ we obtain
\begin{multline*}
R(z) = \frac{\q- zP}{(q-q^{-1}z)} - \frac{\q-z(z-1)}{(q-q^{-1}z)(z+q^{m-n})}
  \vl{}{a,b}{(-1)^{[a][b]}\xi_a\xi_b} 
  {(\r,\ve{a}-\ve{b})}{a}{b}{\bar{a}}{\bar{b}}\\
-\frac{(z-1)}{(q-q^{-1}z)}\left\{ I + (q^{1/2}-q^{-1/2})
  \eso{}{a}{[a]}{a}{a}{a}{a} + \q- \sum_{\ve{a} < \ve{b}} (-1)^{[b]} 
  E^a_b \otimes \hat{\sigma}^b_a \right\}.
\end{multline*} 
 
We comment that although the above derivation only holds for $m>2$, 
the final result holds for all $m$ (see \cite{karen1,karen2}).

\end{section}

\section{The Reshetikhin Twist}

Let $(A, \Delta, R)$ denote a quasi-triangular Hopf superalgebra where $\Delta$ and 
$R$ denote the coproduct and $R$-matrix respectively.  
Consider an element $F \in A \otimes A$ satisfying the properties

\begin{align*}
&(\Delta \otimes I) F = F_{13} F_{23}, \\
&(I \otimes \Delta) F = F_{13} F_{12} ,\\
&F_{12} F_{13} F_{23} = F_{23} F_{13} F_{12}.
\end{align*}

\noindent Then $(A, \Delta^F, R^F)$ is also a quasi-triangular Hopf superalgebra with coproduct and $R$-matrix given by

\begin{equation*}
\Delta^F = F_{12} \Delta F^{-1}_{12}, \qquad R^F = F_{21}RF^{-1}_{21}.
\end{equation*} 

\noindent We refer to F as a {\it twist element}.  In particular, for the case of a 
quantised superalgebra $U_q[g]$ Reshetikhin \cite{reshetikhin} gave 
the example where $F$ is given by

\begin{equation*}
F = \text{exp}\bigl[\sum_{b<c} (h_b \otimes h_c - h_c \otimes h_b) \phi_{bc}
\bigr]
\end{equation*}

\noindent with $\{h_b\}$ the generators of the Cartan subalgebra of $U_q[g]$ and the 
$\phi_{bc},\;  b<c$, arbitrary complex parameters.  

Applying this twist to $\rh(z)$, it is found that both $U_q[osp(m|n)^{(1)}]$ and 
$U_q[gl(m|n)^{(2)}]$ are quasi-triangular Hopf superalgebras with coproduct 
$\Delta^F$ as above and $R$-matrix in the fundamental representation given by

\begin{multline*}
R^F(z) = \frac{\q- zP}{(q-q^{-1}z)} - \frac{\q-z(z-1)}{(q-q^{-1}z)(z-q^{m-n-2})}
  \vl{}{a,b}{(-1)^{[a][b]}\xi_a\xi_b}
  {(\r,\ve{a}-\ve{b})}{a}{b}{\bar{a}}{\bar{b}}\\
-\frac{(z-1)}{(q-q^{-1}z)}\left\{ \biggl(I + (q^{1/2}-q^{-1/2})
  \eso{}{a}{[a]}{a}{a}{a}{a}\biggr)\; \text{exp} \bigl[\sum_{b<c} 
  2(\pi(h_c) \otimes \pi(h_b) - \pi(h_b) \otimes \pi(h_c)) \phi_{bc}\bigr] \right.\\
\left. + \q- \sum_{\ve{a} < \ve{b}} (-1)^{[b]} 
  E^a_b \otimes \hat{\sigma}^b_a \right\}
\end{multline*} 

\noindent for $U_q[osp(m|n)^{(1)}]$ and

\begin{multline*}
R^F(z) = \frac{\q -zP}{(q-q^{-1}z)} - \frac{\q-z(z-1)}{(q-q^{-1}z)(z+q^{m-n})}
  \vl{}{a,b} {(-1)^{[a][b]} \xi_a \xi_b} {(\r,\ve{a}-\ve{b})}{a}{b}{\bar{a}}
  {\bar{b}}\\
-\frac{(z-1)}{(q-q^{-1}z)}\left\{ \biggl( I + (q^{1/2}-q^{-1/2})
  \eso{}{a}{[a]}{a}{a}{a}{a}\biggr) \; \text{exp} \bigl[\sum_{b<c} 
  2(\pi(h_c) \otimes \pi(h_b) - \pi(h_b) \otimes \pi(h_c)) \phi_{bc}\bigr]  \right.\\
 \left. +\q- \sum_{\ve{a} < \ve{b}} (-1)^{[b]} 
  E^a_b \otimes \hat{\sigma}^b_a \right\}
\end{multline*} 
 
\noindent for $U_q[gl(m|n)^{(2)}]$.
In the above formulae the representations $\pi(h_b)$ are given by Table \ref{vector}. 
For both cases we have obtained models with $(l+k)(l+k-1)/2$ continuous variables (the $\phi_{ab}$) and $m+n$
discrete variables (the grading terms $(-1)^{[a]}$: note that there must be an even number of indices 
$a$ for which $[a]=1$, and also that the $\hat{\sigma}_b^a $ explicitly depend on them). 
These variables are in addition to the spectral parameter $z$.
Both models may be considered as generalisations of the Perk--Schultz model.

Independently, similar results have been reported in \cite{marcio}. 
\section*{Acknowledgements} We thank the Australian Research Council for financial support.

\end{document}